\documentclass{webofc}
\usepackage[varg]{txfonts}   
\begin{document}
\title{NIKA2 observations of starless cores in Taurus and Perseus}
%
%

\author{%
  \lastname{C.~Kramer}\inst{\ref{IRAMF}}\fnsep\thanks{\email{kramer@iram.fr}}
  \and  R.~Adam \inst{\ref{OCA}}
  \and  P.~Ade \inst{\ref{Cardiff}}
  \and  H.~Ajeddig \inst{\ref{CEA}}
  \and  P.~Andr\'e \inst{\ref{CEA}}
  \and  E.~Artis \inst{\ref{LPSC},\ref{Garching}}
  \and  H.~Aussel \inst{\ref{CEA}}
  \and  A.~Beelen \inst{\ref{LAM}}
  \and  A.~Beno\^it \inst{\ref{Neel}}
  \and  S.~Berta \inst{\ref{IRAMF}}
  \and  L.~Bing \inst{\ref{LAM}}
  \and  O.~Bourrion \inst{\ref{LPSC}}
  \and  M.~Calvo \inst{\ref{Neel}}
  \and  P.~Caselli \inst{\ref{MPE}}                 
  \and  A.~Catalano \inst{\ref{LPSC}}
  \and  M.~De~Petris \inst{\ref{Roma}}
  \and  F.-X.~D\'esert \inst{\ref{IPAG}}
  \and  S.~Doyle \inst{\ref{Cardiff}}
  \and  E.~F.~C.~Driessen \inst{\ref{IRAMF}}
  \and  G.~Ejlali \inst{\ref{Tehran}}
  \and  A.~Fuente \inst{\ref{CAB}}                  
  \and  A.~Gomez \inst{\ref{CAB}} 
  \and  J.~Goupy \inst{\ref{Neel}}
  \and  C.~Hanser \inst{\ref{LPSC}}
  \and  S.~Katsioli \inst{\ref{Athens_obs}, \ref{Athens_univ}}
  \and  F.~K\'eruzor\'e \inst{\ref{Argonne}}
  \and  B.~Ladjelate \inst{\ref{IRAME}} 
  \and  G.~Lagache \inst{\ref{LAM}}
  \and  S.~Leclercq \inst{\ref{IRAMF}}
  \and  J.-F.~Lestrade \inst{\ref{LERMA}}
  \and  J.~F.~Mac\'ias-P\'erez \inst{\ref{LPSC}}
  \and  S.~C.~Madden \inst{\ref{CEA}}
  \and  A.~Maury \inst{\ref{CEA}}
  \and  P.~Mauskopf \inst{\ref{Cardiff},\ref{Arizona}}
  \and  F.~Mayet \inst{\ref{LPSC}}
  \and  A.~Monfardini \inst{\ref{Neel}}
  \and  A.~Moyer-Anin \inst{\ref{LPSC}}
  \and  M.~Mu\~noz-Echeverr\'ia \inst{\ref{LPSC}}
  \and  D.~Navarro-Almaida \inst{\ref{CEA}}         
  \and  L.~Perotto \inst{\ref{LPSC}}
  \and  G.~Pisano \inst{\ref{Roma}}
  \and  N.~Ponthieu \inst{\ref{IPAG}}
  \and  V.~Rev\'eret \inst{\ref{CEA}}
  \and  A.~J.~Rigby \inst{\ref{Leeds}}
  \and  A.~Ritacco \inst{\ref{INAF}, \ref{ENS}}
  \and  C.~Romero \inst{\ref{Pennsylvanie}}
  \and  H.~Roussel \inst{\ref{IAP}}
  \and  F.~Ruppin \inst{\ref{IP2I}}
  \and  K.~Schuster \inst{\ref{IRAMF}}
  \and  A.~Sievers \inst{\ref{IRAME}}
  \and  C.~Tucker \inst{\ref{Cardiff}}
  \and  R.~Zylka \inst{\ref{IRAMF}}
}
\institute{
  Institut de RadioAstronomie Millim\'etrique (IRAM), 38406 Saint Martin d'H\`eres, France
  \label{IRAMF}
  \and
  Universit\'e C\^ote d'Azur, Observatoire de la C\^ote d'Azur, CNRS, Laboratoire Lagrange, France 
  \label{OCA}
  \and
  School of Physics and Astronomy, Cardiff University, CF24 3AA, UK
  \label{Cardiff}
  \and
  Universit\'e Paris-Saclay, Université Paris Cité, CEA, CNRS, AIM, 91191, Gif-sur-Yvette, France
  \label{CEA}
  \and
  Universit\'e Grenoble Alpes, CNRS, Grenoble INP, LPSC-IN2P3, 38000 Grenoble, France
  \label{LPSC}
  \and	
  Max Planck Institute for Extraterrestrial Physics, 85748 Garching, Germany
  \label{Garching}
  \and
  Aix Marseille Univ, CNRS, CNES, LAM, Marseille, France
  \label{LAM}
  \and
  Universit\'e Grenoble Alpes, CNRS, Institut N\'eel, France
  \label{Neel}
  \and 
  Dipartimento di Fisica, Sapienza Universit\`a di Roma, I-00185 Roma, Italy
  \label{Roma}
  \and
  Univ. Grenoble Alpes, CNRS, IPAG, 38000 Grenoble, France
  \label{IPAG}
  \and
  Institute for Research in Fundamental Sciences (IPM), Larak Garden, 19395-5531 Tehran, Iran
  \label{Tehran}
  \and
  Centro de Astrobiolog\'ia (CSIC-INTA), Torrej\'on de Ardoz, 28850 Madrid, Spain
  \label{CAB}
  \and
  National Observatory of Athens, IAASARS, GR-15236, Athens, Greece
  \label{Athens_obs}
  \and
  Faculty of Physics, University of Athens, GR-15784 Zografos, Athens, Greece
  \label{Athens_univ}
  \and
  High Energy Physics Division, Argonne National Laboratory, Lemont, IL 60439, USA
  \label{Argonne}
  \and  
  Instituto de Radioastronom\'ia Milim\'etrica (IRAM), Granada, Spain
  \label{IRAME}
  \and
  LERMA, Observatoire de Paris, PSL Research Univ., CNRS, Sorbonne Univ., UPMC, 75014 Paris, France  
  \label{LERMA}
  \and
  School of Earth \& Space and Department of Physics, Arizona State University, AZ 85287, USA
  \label{Arizona}
  \and
  School of Physics and Astronomy, University of Leeds, Leeds LS2 9JT, UK
  \label{Leeds}
  \and
  INAF-Osservatorio Astronomico di Cagliari, 09047 Selargius, Italy
  \label{INAF}
  \and 
  LPENS, ENS, PSL Research Univ., CNRS, Sorbonne Univ., Universit\'e de Paris, 75005 Paris, France 
  \label{ENS}
  \and  
  Department of Physics and Astronomy, University of Pennsylvania, PA 19104, USA
  \label{Pennsylvanie}
  \and
  Institut d'Astrophysique de Paris, CNRS (UMR7095), 75014 Paris, France
  \label{IAP}
  \and
  University of Lyon, UCB Lyon 1, CNRS/IN2P3, IP2I, 69622 Villeurbanne, France
  \label{IP2I}
}
       
\abstract{Dusty starless cores play an important role in regulating
  the initial phases of the formation of stars and planets. In their
  interiors, dust grains coagulate and ice mantles form, thereby
  changing the millimeter emissivities and hence the ability to
  cool. We mapped four regions with more than a dozen cores in the
  nearby Galactic filaments of Taurus and Perseus using the NIKA2
  camera at the IRAM 30-meter telescope.  Combining the 1\,mm to 
  2\,mm flux ratio maps with dust temperature maps from Herschel
  allowed to create maps of the dust emissivity index $\beta_{1,2}$ at
  resolutions of 2430 and 5600\,a.u. in Taurus and Perseus,
  respectively. Here, we study the variation with total column
  densities and environment. $\beta_{1,2}$ values at the core centers
  ($A_V=12-19$\,mag) vary significantly between $\sim1.1$ and
  $2.3$. Several cores show a strong rise of $\beta_{1,2}$ from the
  outskirts at $\sim4$\,mag to the peaks of optical extinctions,
  consistent with the predictions of grain models and the gradual
  build-up of ice mantles on coagulated grains in the dense interiors
  of starless cores.  }
\maketitle
\section{Cores in Taurus and Perseus}
\label{sec-introduction}

Herschel images of giant molecular clouds and dark cloud complexes
have revealed large networks of filamentary structures where stars are
born \cite{Andre2010}.  Filaments precede the onset of most star
formation, funneling interstellar gas and dust into increasingly
denser concentrations that will contract and fragment leading to
gravitationally bound starless cores that will eventually form
stars. In the cold dense interiors most molecules freeze-out
\cite{Kramer1999, Caselli2022}, covering the grains with icy mantles, which make
them sticky, favoring grain coagulation, altering the grain size
distribution, and their emissivity \cite{Ormel2011}.
%
%
%
%
%


The dual-band camera NIKA2 \cite{Perotto2020} at the IRAM 30-meter
telescope was used to map four regions in the nearby molecular
filaments of Taurus and Perseus simultaneously at 1 and 2\,mm
wavelengths.  The regions were selected from a survey of gas phase
abundances in 29 starless cores covering different star formation
activity, {\tt GEMS}\footnote{Gas phase Elemental abundances in
  Molecular clouds, https://www.oan.es/gems/doku.php, PI: A.~Fuente},
a large program using the EMIR eight-band heterodyne receiver at the
30-meter telescope \cite{Fuente2019, Navarro-Almaida2023}.  The gas
phase abundances of Carbon, Oxygen, Nitrogen, and Sulphur were
measured together with the ionization fraction $X$(e$^-$) by spatially
resolved cuts of molecular key tracers through the cores.


The NIKA2 cores lie in Taurus and Perseus, two nearby filamentary
molecular cloud complexes. The Taurus clouds show isolated low-mass
star formation and lie at a solar distance of only 135\,pc
allowing us to reach resolutions of 2430\,a.u. ($18''$)
with NIKA2. The Perseus region exhibits the formation of stellar clusters
and lies at a distance of 310\,pc.
Their cores sample a range of peak column densities parametrized by
optical extinctions (12 to 19\,mag), dust temperatures averaged along
the lines of sight of 11 to 19\,K, volume densities of $\sim10^4$ to
$10^5$\,cm$^{-3}$ (Table\,\ref{tab-cores}), and gas phase fractional
abundances. The cores are at different evolutionary stages and are
located in different environments, from low-mass to massive
star-forming regions. Different star formation activities translate
into different UV fluxes and dust temperatures. Higher dust
temperatures ($\sim$20 - 30\,K) hinder the freeze-out of volatile
molecules.

\begin{table}[ht!]
\caption{Properties of the cores observed with NIKA2: region, number
  of selected cores, range of peak optical extinctions, dust
  temperatures \cite{SinghMartin2022}, volume densities
  \cite{Fuente2019, RodriguezBaras2021, Navarro-Almaida2021}, dust
  emissivities $\beta_{1,2}$ (this study, Figs.\,\ref{fig-beta-maps},
  \ref{fig-beta-histograms}). The last column lists the star formation
  activity of the region. } \centering
\begin{tabular}{lrrcccl}\\ \hline \hline
  Region          &  \# & $A_V$ & $T_d$ & $n$(H$_2$) & $\beta_{1,2}$ & SF activity    \\
                  &     & (mag) & (K)   & ($10^4$\,cm$^{-3}$)      &                \\ \hline \hline
  %
  TMC1/Taurus     &   2 &  17-18 &  12-13 &  2-5   & 1.8-1.9       & Low            \\
  B213/Taurus     &   2 &  15-17 &  11-12 &  5-6   & 1.8-2.3       & Low            \\
  NGC1333/Perseus &   6 &  12-17 &  12-19 &  5-11  & 1.1-1.9       & Intermediate   \\
  IC348/Perseus   &   2 &  17-19 &  16-17 &  5-9   & 1.7-1.8       & High           \\ \hline \hline
\end{tabular}
\label{tab-cores}
\end{table}

\begin{figure}[h]
\centering
\includegraphics[scale=0.6]{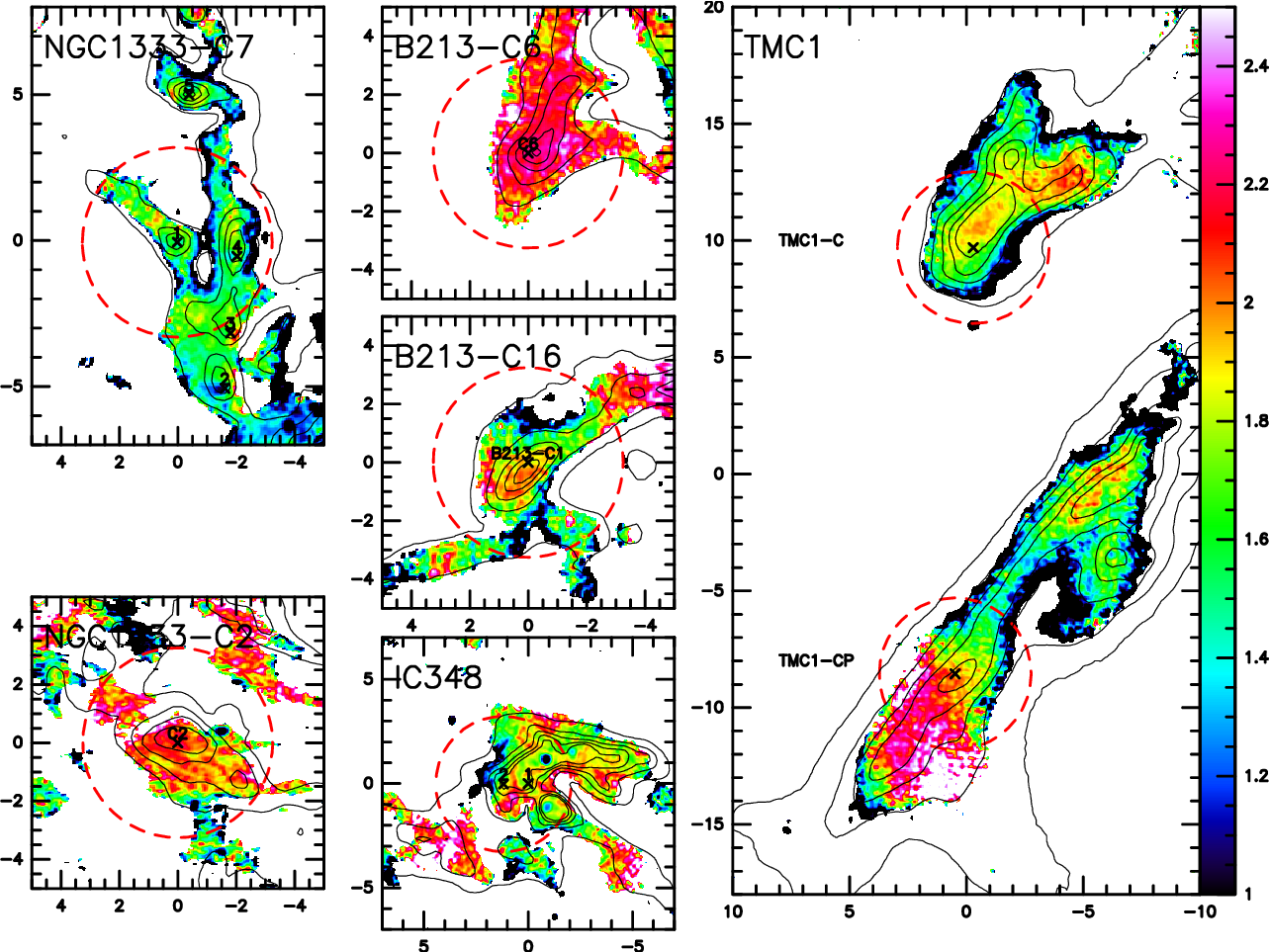}
\caption{Maps of the dust emissivity index $\beta_{1,2}$ at a
  resolution of $18''$ for the regions mapped with NIKA2. Positions
  with fluxes below $6\sigma$ were masked. The range in $\beta_{1,2}$
  is provided in the colour bar. Contours of optical extinctions are
  repeated at levels of 4, 7, 10, 13, 16\,mag. Coordinates are
  $\Delta$R.A. and $\Delta$Dec. offsets in arcminutes. The NIKA2
  field-of-view of $6.5'$ is marked by red dashed circles. Black
  crosses mark the positions of selected cores
  (cf.\,Fig.\,\ref{fig-beta-histograms}). }
\label{fig-beta-maps}
\end{figure}

\section{NIKA2 observations}
\label{sec-nika-2-observations}

The NIKA2 observations comprise 67.3\,hours of telescope time. They
were conducted in the years 2019 to 2022 in the framework of a series
of open time proposals (027-19, 128-19, 006-20, 096-20, 010-21,
110-21, 008-22 (PI C.~Kramer). The total observed area covers more
than 1000\,arcmin$^2$ (Fig.\,\ref{fig-beta-maps}). The average
$1\sigma$ rms noise values are 0.3\,mJy/beam at 2\,mm and
0.9\,mJy/beam at 1\,mm. The NIKA2 observations were reduced adapting
the {\tt piic/gildas} pipeline provided by IRAM
\cite{BertaZylka2023}. The average point source flux uncertainties
during all runs are 6\% at 2\,mm and 8\% at 1\,mm, resulting in
uncertainties of the ratios better than 10\,\%.  The relative
calibration benefits from simultaneously observing under the same
conditions and in particular through the same atmosphere. The NIKA2
observations were reduced adapting the {\tt piic/gildas} pipeline
provided by IRAM \cite{BertaZylka2023}.

\begin{figure}[h]
\centering
\includegraphics[scale=0.6]{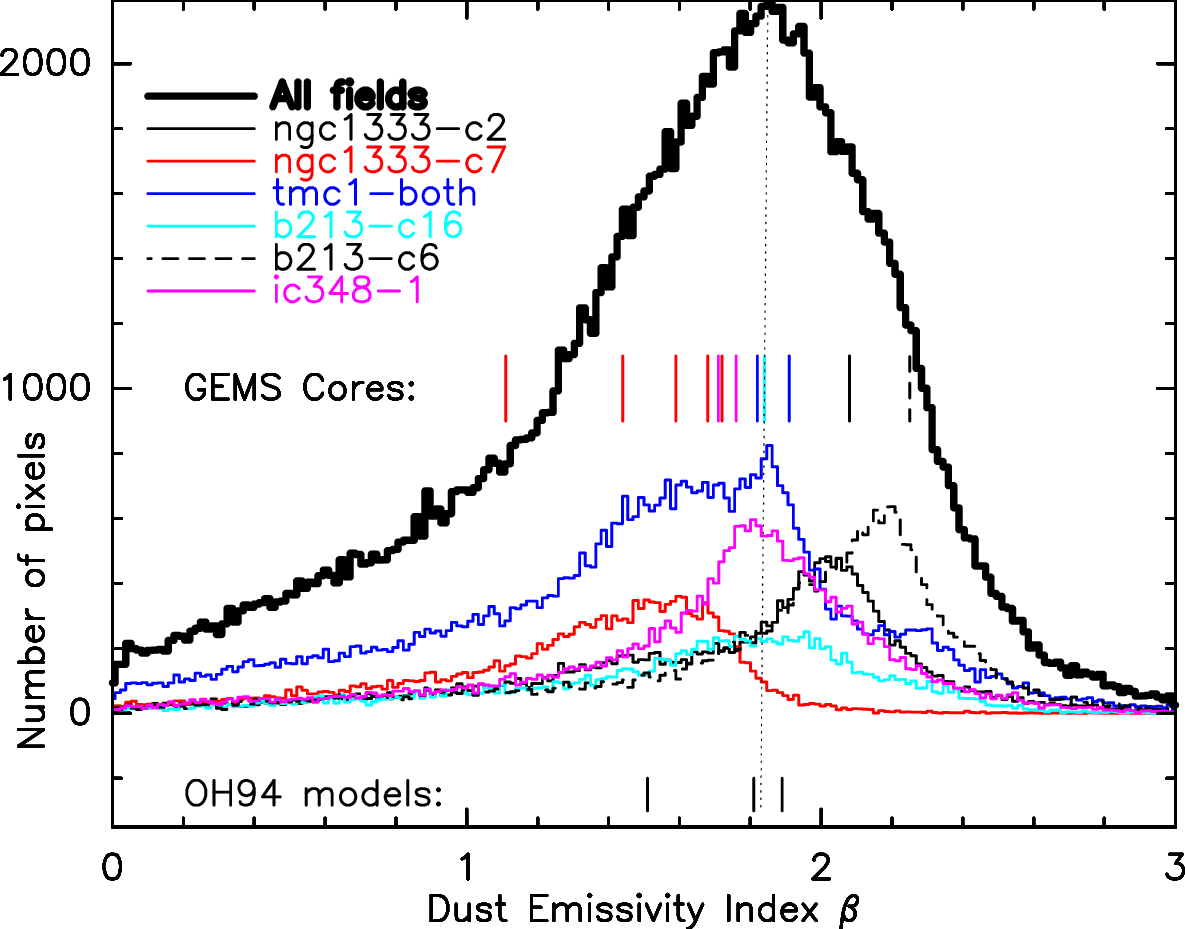}
\caption{Number of map pixels per interval of the dust emissivity
  index $\beta_{1,2}$.  The black histogram with thick lines shows the
  cumulative distribution for all regions with NIKA2 shown in
  Fig.\,\ref{fig-beta-maps}, while the colored histograms show the
  distributions for each of the observed regions.  The observed GEMS
  cores are marked by short, colored, vertical lines. At the bottom,
  black vertical lines show that indices of OH94 grain models
  \cite{OH94} for coagulated grains at $n_{\rm H}=10^5$\,cm$^{-3}$ and
  varying the thickness of ice mantles
  (cf. Fig.\,\ref{fig-oh94-models}, Right). }
\label{fig-beta-histograms} 
\end{figure}
%

\section{The dust emissivity index}

The observed intensities at a given frequency $\nu$ are given by
$I_\nu=\tau_\nu B_\nu(T_d)=\kappa_\nu\Sigma B_\nu(T_d)$ with the
opacity $\tau_\nu$, the Planck function $B_\nu$, the dust temperature
$T_d$, the emissivity cross section per gram of dust and gas
$\kappa_\nu$, and the gas surface density $\Sigma$. At the NIKA2
wavelengths, the assumption of optically thin emission is well
justified. The emissivity cross section is often parametrized by
$\kappa_\nu=\kappa_0(\nu/\nu_0)^\beta$ introducing the long wavelength
dust emissivity slope $\beta$.
%
%
As the Rayleigh-Jeans approximation is no longer valid at the low
temperatures encountered in prestellar cores, the 1\,mm/2\,mm flux
ratio $R_{1,2}$ is weakly dependent on the dust temperature: $R_{1,2}
= B_1(T_d)/B_2(T_d)(\nu_1/\nu_2)^{\beta_{1,2}}$.  Fits of modified
black-bodies to PACS and SPIRE 160, 250, 350, 500$\mu$m data from the
Herschel Gould Belt Survey (HGBS) \cite{Andre2010} were used by
\cite{SinghMartin2022} to create maps of dust temperature and optical
extinction at $36.4''$ resolution. These maps were used here.

The NIKA2 1\,mm maps were convolved with a Gaussian kernel to the
resolution at 2\,mm, $18''$, and the two maps were registered onto the
same grid to create maps of the flux ratio $R_{1,2}$ and of the dust
emissivity index $\beta_{1,2}(R_{1,2}, T_d)$
(Fig.\,\ref{fig-beta-maps}). Positions with fluxes below $6\sigma$
were masked. In future work, we plan for a more thorough analysis of
errors, taking into account the work by \cite{Rigby2018} who studied
dust emissivity variations of two star-forming clouds observed with
NIKA.
In the present work, the $\beta_{1,2}$ maps show significant
variations between $\sim1.2$ and 2.4. This spread is seen in several
but not all of the individual maps (Fig.\,\ref{fig-beta-maps}). The
core centers where the optical extinctions peak show a similar spread
of $\beta_{1,2}$ values (Fig.\,\ref{fig-beta-histograms}). The
variations of $\beta_{1,2}$ observed by us are well above the
calibration errors and indicate systematic variations of the grain
population emitting at mm wavelengths. Similar $\beta$ values have
been found in these and other cores of Taurus using also other
millimeter/submillimeter cameras
\cite{Nguyen2023, Navarro-Almaida2023, Scibelli2023, Bracco2017, Chacon-Tanarro2019}.

\begin{figure}[h] 
\centering
\includegraphics[scale=0.48]{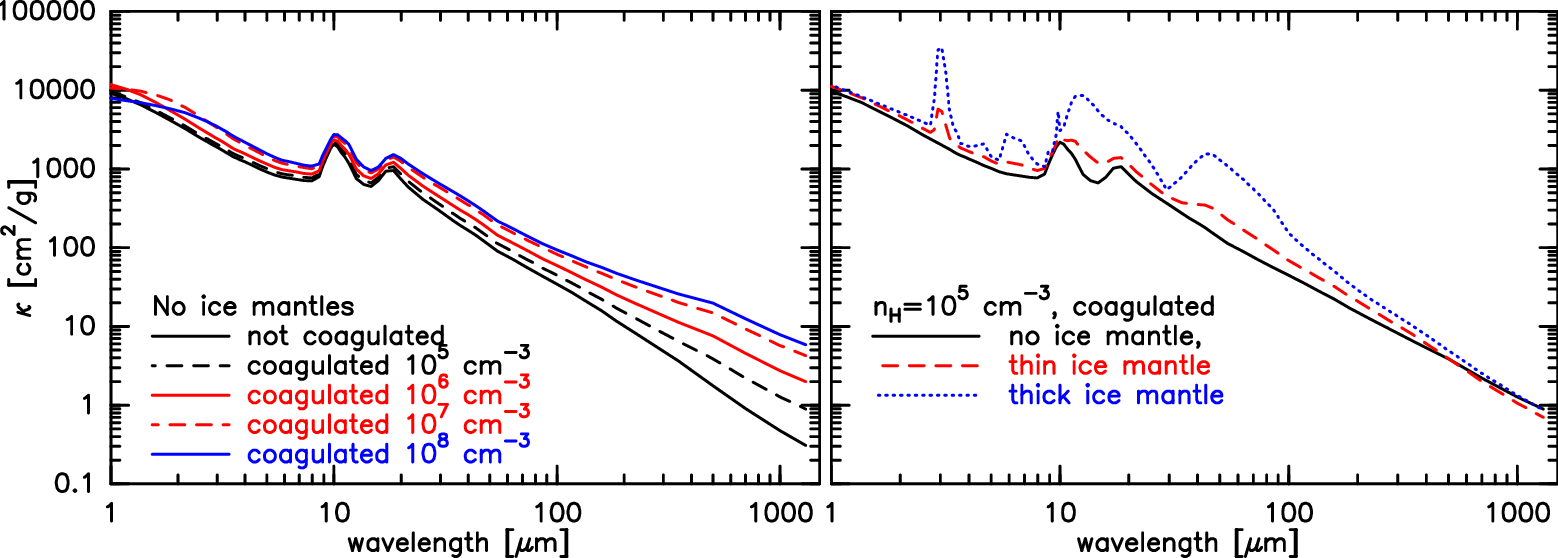}
\caption{Modelled dust opacities \cite{OH94}. {\bf Left:} Bare grains
  without ice mantles, not coagulated and coagulated with increasing
  hydrogen densities $n_{\rm H}$. $\beta_{1.3,0.7}$ decreases from 1.74 to 1.23.
  {\bf Right:} Coagulated grains with increasing ice mantle thickness
  at a constant hydrogen density. NIR ice features start to show-up
  and the dust emissivity index $\beta_{1.3,0.7}$ steepens from 1.51 to 1.89. }
\label{fig-oh94-models}       
\end{figure}

\section{Dust opacity models}

To understand these indices and their variations better, we compare
with grain models. Figure\,\ref{fig-oh94-models} shows dust opacity
spectra modelled by \cite{OH94} for proto-stellar cores. These models
assume dust grains consisting of a mixture of silicates and amorphous
carbon with different levels of coagulation and ice layer coverage
on the agglomerates. Starting with the Mathis-Rumpl- Nordsieck
(MRN) grain size distribution \cite{Mathis1977}, a coagulation period of
$10^5$\,years is simulated.
NIR/MIR wavelengths are marked by narrow silicate and ice features
while submm/mm wavelengths exhibit marked variations of the slope,
i.e. of the dust emissivity index $\beta$.  For bare grains, the
$\beta_{1.3,0.7}$ index drops from 1.74 to 1.23 with increasing density, when
coagulation leads to increasingly large grains
(Fig.\,\ref{fig-oh94-models}, Left). On the other hand, $\beta_{1.3,0.7}$
increases with increasing thickness of the ice mantles
(Fig.\,\ref{fig-oh94-models}, Right) from 1.5 to 1.89 for coagulated
grains and a fixed density of $n_{\rm H}=10^5$\,cm$^{-3}$.

\section{Conclusions}

Most observations of $\beta_{1,2}$ are consistent with the gradual
build-up of ice mantles in the interiors of
pre-stellar cores at the volume densities of $\sim10^5$\,cm$^{-3}$
(Table\,\ref{tab-cores}) as predicted by grain models \cite{OH94}
(cf.\,Fig.\,\ref{fig-oh94-models}, Right). In particular the radial
drop of $\beta_{1,2}$ in TMC1-C fits this picture. Here, $\beta_{1,2}$ drops
from $\sim2.0$ at the $A_V$ peak with $\sim16$\,mag to the outskirts
at $\sim4$\,mag with $\beta_{1,2}\sim1.1$. The region B213-C16 also
fits this picture.
The filamentary structure of NGC1333-C7 exhibits an almost constant
$\beta_{1,2}\sim1.6$ with the sole exception of core \#2 exhibiting a
much lower value of $\sim1.0$. Interestingly, core \#2 is a Class I
protostellar object and its Deuteration, measured by the ratio of
DCN/H$^{13}$CN abundances, is significantly lower than for the other
cores of this region \cite{Navarro-Almaida2023}. Its low $\beta_{1,2}$ value
indicates bare grains, devoid of ice mantles, consistent with a more
evolved state. The other cores of this filament are less evolved
objects \cite{Hacar2017} in which CO freezes-out leading to the
formation of ice mantles, and driving deuteration.
The $\beta_{1,2}$ map of B213-C6 differs from the other cores. It
exhibits fairly constant and high values of $\sim2.2$, hardly varying
with $A_V$.  The parameter space covered by the OH94 models does
however not reproduce such high $\beta_{1,2}$ values of $>2.0$. See
also the discussion in \cite{Scibelli2023}. Alternative models to
explain such high values are discussed in \cite{Lippok2016}.

\section*{Ackowledgements}
\small{We would like to thank the IRAM staff for their support during the observation campaigns. The NIKA2 dilution cryostat has been designed and built at the Institut N\'eel. In particular, we acknowledge the crucial contribution of the Cryogenics Group, and in particular Gregory Garde, Henri Rodenas, Jean-Paul Leggeri, Philippe Camus. This work has been partially funded by the Foundation Nanoscience Grenoble and the LabEx FOCUS ANR-11-LABX-0013. This work is supported by the French National Research Agency under the contracts "MKIDS", "NIKA" and ANR-15-CE31-0017 and in the framework of the "Investissements d’avenir” program (ANR-15-IDEX-02). This work has benefited from the support of the European Research Council Advanced Grant ORISTARS under the European Union's Seventh Framework Programme (Grant Agreement no. 291294). A. R. acknowledges financial support from the Italian Ministry of University and Research - Project Proposal CIR01$\_00010$. S. K. acknowledges support provided by the Hellenic Foundation for Research and Innovation (HFRI) under the 3rd Call for HFRI PhD Fellowships (Fellowship Number: 5357). 
}


\begin{thebibliography}{}
%
%

%
\bibitem{Andre2010}
  P.~Andr\'e {\it et al.}, Astron. Astrophys. \textbf{518}, 102 (2010)
\bibitem{BertaZylka2023}
  S.~Berta \& R.~Zylka, {\tt piic} user manual, IRAM report (2023)
\bibitem{Bracco2017}
  A.~Bracco {\it et al.}, Astron. Astrophys. \textbf{604}, A52 (2017)
\bibitem{Caselli2022}
  P.~Caselli {\it al.}, Astron. Astrophys. \textbf{929}, 13 (2022)
\bibitem{Chacon-Tanarro2019}
  A.~Chac\'on-Tanarro {\it et al.}, Astron. Astrophys. \textbf{623}, 118 (2019) 
\bibitem{Fuente2019}
  A.~Fuente {\it et al.}, Astron. Astrophys. \textbf{624}, 105 (2019)
\bibitem{Hacar2017}
  A.~Hacar {\it et al.}, Astron. Astrophys. \textbf{606}, 123 (2017)
\bibitem{Kramer1999}
  C.~Kramer {\it et al.}, Astron. Astrophys. \textbf{342}, 257 (1999)
\bibitem{Lippok2016}
  N.~Lippok {\it et al.}, Astron. Astrophys. \textbf{592}, A61 (2016)
\bibitem{Mathis1977}
  J.S.~Mathis {\it et al.}, ApJ \textbf{217}, 425 (1977)
\bibitem{Navarro-Almaida2021}
  D.~Navarro-Almaida {\it et al.}, Astron. Astrophys. \textbf{653}, A15 (2021)
\bibitem{Navarro-Almaida2023}
  D.~Navarro-Almaida {\it et al.}, Astron. Astrophys. \textbf{670}, A110 (2023)
\bibitem{Nguyen2023}
  Q.~Nguyen Luong {\it et al.}, these proceedings (2023)
\bibitem{Ormel2011}
  C.W.~Ormel {\it et al.}, Astron. Astrophys. \textbf{532}, A43 (2011)
\bibitem{OH94}
  V.~Ossenkopf \& T.~Henning, Astron. Astrophys. \textbf{291}, 943 (1994)
\bibitem{Perotto2020}
  L.~Perotto, N.~Ponthieu, J.-F.~Mac\'ias-P\'erez, {\it et al.},
  Astron.\ Astrophys.\  {\bf 637}, A71 (2020)
\bibitem{Rigby2018}
  A.J.~Rigby {\it et al.}, Astron. Astrophys. \textbf{615}, A18 (2018)
\bibitem{RodriguezBaras2021}
  M.~Rodriguez-Baras {\it et al.}, Astron. Astrophys. \textbf{648}, A120 (2021)
\bibitem{Schnee2014}
  S.~Schnee {\it et al.}, MNRAS, \textbf{444}, 2303 (2014)
\bibitem{Scibelli2023}
  S.~Scibelli {\it et al.}, MNRAS, \textbf{521}, 4579 (2023)
\bibitem{SinghMartin2022}
  A.~Singh \& P.G.~Martin {\it et al.}, ApJ \textbf{941}, 135 (2022)
\end{thebibliography}
\end{document}